\documentclass[aps,pra,twocolumn,showpacs,amssymb,amsmath,amsfonts,superscriptaddress,floatfix,nofootinbib,groupedaddress]{revtex4}
\usepackage{graphicx,graphics,color,times,bm,bbm}
\newcommand{\ket}[1]{|#1\rangle}
\newcommand{\bra}[1]{\langle #1|}
\newcommand{\braket}[2]{\langle #1|#2\rangle}

\begin{document}

\title{Bound entanglement in quantum phase transitions}

\author{S. Baghbanzadeh}
\affiliation{Department of Physics, Sharif University of Technology, Post Office Box 11155-9161, Tehran, Iran}
\author{S. Alipour}
\affiliation{Department of Physics, Sharif University of Technology, Post Office Box 11155-9161, Tehran, Iran}
\author{A. T. Rezakhani}
\affiliation{Department of Chemistry and Center for Quantum Information Science and Technology, University of Southern California, Los Angeles, California 90089, USA}

\begin{abstract}
We investigate quantum phase transitions in which a change in the type of entanglement from bound entanglement to either free entanglement or separability may occur. In particular, we present a theoretical method to construct a class of quantum spin-chain Hamiltonians that exhibit this type of quantum criticality. Given parameter-dependent two-site reduced density matrices (with prescribed entanglement properties), we lay out a reverse construction for a compatible pure state for the whole system, as well as a class of Hamiltonians for which this pure state is a ground state. This construction is illustrated through several examples.
\end{abstract}
\maketitle

\section{Introduction}
\label{intro}

Entanglement, a key concept in quantum-information science \cite{Nielsen:book}, has also been playing a pivotal role in the study of quantum (or, generally, nonclassical) correlations and related consequences in many-body systems \cite{RMPent}, especially quantum phase transitions (QPTs) induced by varying parameters of system Hamiltonian at zero temperature \cite{sachdev} (see also Refs. \cite{qpt-gm} for geometric approaches to QPTs.). Accordingly, many efforts have been made to find the possible relation between QPT in a system and nonanalyticities of, e.g., pairwise entanglement of its ground state (GS) \cite{osterloh}. For example, under some fairly general conditions, a classification of QPTs in terms of pairwise entanglement has been provided in Ref.~\cite{wsl-prl}. Besides, through density functional theoretical arguments, it has been shown that by an appropriate choice of entanglement measure (either bi- or multipartite), one can study physical properties of a system near its critical points \cite{wsl-pra}. It is thus evident that having a sufficiently strong entanglement measure which is able to detect entanglement regions of the GS (in terms of the Hamiltonian parameters) in a more subtle way is of paramount importance. The difficulty of the issue, i.e., quantifying quantum correlations, is mostly due to the existence of various types of entanglement in quantum many-body systems, such as free entanglement (FE), bound entanglement (BE) \cite{BE}, and pairwise or multipartite entanglement \cite{Horodeckis}, featuring also new phases for quantum matter (e.g., topological order \cite{top}). Here, however, we shall concentrate only on BE.

Simply speaking, bound entangled states, in contrast to free entangled states, are those states which are not directly useful for quantum-information tasks such as teleportation \cite{tlpdc}. Because of the positive partial transpose (PPT) property of bound entangled states (in the bipartite case) \cite{BE}, their entanglement cannot be distilled by any local operation and classical communication; for a general review, see, e.g., Ref.~\cite{Alber:book}. The possibility of the existence of non-PPT BE, however, is still an open question \cite{npt}. Nevertheless, bound entangled states are still useful because, for example, cryptographically secure keys can be constructed from them \cite{Hor-key}, they can enhance the fidelity of conclusive teleportation---hence, the teleportation power---of another bipartite state \cite{contlp}, they manifest irreversibility in the entanglement formation-distillation cycle (hence, also of interest in thermodynamics) \cite{irrev}, and they have applications in channel discrimination \cite{chdis}. It is, therefore, intriguing to see how or whether BE can appear in quantum many-body systems. Recently, it has been shown that a thermal environment can naturally induce the thermal transition from FE to BE in quantum many-body systems \cite{tem}. Additionally, in a simple $XY$ spin chain, varying the magnetic field can generate bound entangled GSs \cite{patane}. In a recent experiment \cite{Expr}, an optical four-qubit state with BE has been realized and then characterized fully.

There is vast literature on QPTs induced by a transition from separability to FE or a transition in the FE region (in the
sense of bipartite reduced density matrices of the GSs). In this work, in contrast, we are interested in QPTs in which the transition happens because of a change in the type of entanglement, specifically, from FE to BE (or conversely), or from separability to BE (or conversely). This is particularly interesting, because one of the vastly used measures of bipartite entanglement in the context of QPTs, negativity \cite{neg}, is inherently unable of detecting BE, hence it would fail to signal a corresponding criticality. There are, however, some other computable (but not necessarily conclusive) inseparability tests for recognizing BE, based on the cross norm \cite{realign} and permutation \cite{permu} criteria. In the case of quantum spin chains, a complete set of spin squeezing inequalities has also been introduced which can be used for experimentally
detecting BE of many-body thermal states \cite{squ}. More appealing and arguably a stronger identifier than entanglement
measures for detecting QPTs is provided by the ground-state fidelity (GSF) \cite{GSF} or its second derivative \cite{FS}, fidelity susceptibility (FS). GSF and FS have been successfully used in the context of QPTs in correlated quantum many-body systems \cite{Gu-review}.

We address the types of QPTs we are interested in through a reverse-construction method.
Given a many-body density matrix $\varrho$, we construct a spin-one chain Hamiltonian $H$ whose GS has $\varrho$ as the reduced density matrix
of specific adjacent sites. Thereby, by tuning the Hamiltonian parameters in some suitable intervals, we will have GSs representing QPTs accompanied by the transition of their entanglement type from BE to FE or separability. The technique is in principle relatively general and is illustrated by three examples. We should remark that we are not aiming to construct physically realizable Hamiltonians. Thus, here we are not concerned with the possible appearance of many-body (three-body or higher order) interactions in the constructed Hamiltonians. This, although might seem to restrict the realistic applicability of the approach, is not a fundamental issue. There exists a powerful perturbative (``gadget") machinery to construct arbitrary $k$-body (effective) interactions from two-body Hamiltonians \cite{gadget}.

\section{Construction of the Hamiltonian}
\label{constH}

Consider a closed chain of $2N$ identical quantum systems, e.g., spins, each of which has a $d$-dimensional Hilbert space. Let us assume that we are given a global state $|\Psi\rangle$ whose two-site reduced density matrices (TSRDMs) have some preset properties. For example, for odd 
$i$, $\mathrm{Tr}_{\overline{i,i+1}}\bigl[|\Psi\rangle_{1,\ldots,2N}\langle\Psi| \bigr]$ must be compatible with an already given $\varrho_{i,i+1}$ 
, where $\overline{i,i+1}=\{1,\ldots,2N\}-\{i,i+1\}$ (we denote the TSRDMs for even $i$ by $\varrho'$).
For simplicity, let us suppose that there is a symmetry by which all $\varrho_{i,i+1}$ ($\varrho'_{i,i+1}$) are the same for all odd (even) $i$. That is, the matrix form of $\varrho_{1,2}$ ($\varrho'_{2,3}$) can be considered as a representative of all $\varrho_{i,i+1}$ ($\varrho'_{i,i+1}$) with odd (even) $i$. We also assume that
\begin{eqnarray}
&\hskip-4mm \varrho_{i,i+1} = \sum_{\imath=1}^K \lambda_{\imath} |v_{\imath} \rangle_{i,i+1} \langle v_{\imath}| + 0 \times \sum_{\jmath=1}^{d^2-K} |w_{\jmath}\rangle_{i,i+1} \langle w_{\jmath}|,
\nonumber\\
&\hskip-4mm \varrho'_{i,i+1} = \sum_{\imath=1}^{K'} \lambda'_{\imath} |v'_{\imath} \rangle_{i,i+1} \langle v'_{\imath}| + 0 \times \sum_{\jmath=1}^{d^2-K'} |w'_{\jmath}\rangle_{i,i+1} \langle w'_{\jmath}|,\nonumber\\
\label{rho'ii1}
\end{eqnarray}
where $0<\lambda_{\imath}\leq 1$ for all $\imath$, $\langle v_{\imath}|v_{\imath'}\rangle = \langle v'_{\imath}|v'_{\imath'}\rangle= \delta_{\imath\imath'}$, $\langle w_{\jmath}|w_{\jmath'}\rangle = \langle w'_{\jmath}|w'_{\jmath'}\rangle = \delta_{\jmath\jmath'}$, and $\langle v_{\imath}|w_{\jmath}\rangle = \langle v'_{\imath}|w'_{\jmath}\rangle =0$.
That is, Eqs.~(\ref{rho'ii1}) are the spectral representations of $\varrho_{i,i+1}$ and $\varrho'_{i,i+1}$ (including degeneracies), where the vectors $\{|w_{\jmath}\rangle\}_{\jmath=1}^{d^2-K}$ and $\{|w'_{\jmath}\rangle\}_{\jmath=1}^{d^2-K'}$ constitute their null eigenspaces.

Our goal is to construct (or, exactly speaking,  reverse construct) a corresponding Hamiltonian $H$ for which this $|\Psi\rangle$ is a GS.
We follow similar steps as in Ref.~\cite{ABK}. In the following, we restrict ourselves only to positive semidefinite Hamiltonians. Thus, for any arbitrary $|\Psi\rangle$, we have $\langle\Psi|H|\Psi\rangle \geq E_0\geq0$, so that if for a $|\Psi\rangle$, $\langle\Psi|H|\Psi\rangle=0$ this is an eigenvector of $H$ with zero eigenvalue, namely, the GS of the Hamiltonian. Moreover, we are interested in nearest-neighbor two-body (or at most three-body) interactions. That is, the sought-after $H$ has the following general form:
\begin{eqnarray}
& H = \sum_{i~\text{odd}} H_{i,i+1} + \sum_{i~\text{even}} H'_{i,i+1}.
\label{newH}
\end{eqnarray}
It then can be shown that
\begin{eqnarray}
&\hskip-5mm\langle\Psi|H|\Psi\rangle = \sum_{i~\text{odd}} \mathrm{Tr}[H_{i,i+1} \varrho_{i,i+1}] + \sum_{i~\text{even}} \mathrm{Tr}[H'_{i,i+1} \varrho'_{i,i+1}].\nonumber\\
\label{php}
\end{eqnarray}
Thus, if for odd (even) $i$ we construct their corresponding $H_{i,i+1}$ ($H'_{i,i+1}$) from the null eigenvectors (or any combination or subset of the vectors in the null eigenspaces) $\{|w_{\jmath}\rangle\}$ ($\{|w'_{\jmath}\rangle\}$), as
\begin{eqnarray}
& H_{i,i+1} = \sum_{\jmath\in\mathcal{J}} h^{[i]}_{\jmath} |w_{\jmath}\rangle_{i, i+1} \langle w_{\jmath}|
\label{Hii1-f}
\end{eqnarray}
(and similarly for $H'_{i,i+1}$), where $\mathcal{J}\subseteq \{1,\ldots,d^2-K\}$ ($\mathcal{J}'\subseteq \{1,\ldots,d^2-K'\}$) and $h^{[i]}_{\jmath}$ (${h'}^{[i]}_{\jmath}$) are arbitrary nonnegative numbers, then all the terms in the summations of Eq.~(\ref{php}) would vanish. This of course is a sufficient condition, not always necessary. Overall, with this choice for $H$, we have $\langle \Psi|H|\Psi\rangle=0$, implying that $|\Psi\rangle$ is a GS.

Put briefly, then, the construction boils down to finding the null eigenspaces of $\varrho_{1,2}$ and $\varrho_{2,3}$. As long as, for the given $|\Psi\rangle$, the nullities of these density matrices are nonzero, one can find a nearest-neighbor two-body Hamiltonian describing the interactions on the chain. Otherwise, the above recipe fails to provide coupling between all the neighboring particles on the chain. This, however, does not necessarily imply nonexistence of other compatible two-body Hamiltonians. To remedy the issue with our construction, given that $|\Psi\rangle$ is still the state with the desired properties, a possible way is to consider higher order interactions, e.g., three body couplings. We should consider $\varrho_{i,i+1,i+2}$, or in general $\varrho_{i,i+1,\ldots,i+L}$ (for some $L$), rather than $\varrho_{i,i+1}$ (for odd $i$). Now if we repeat the Hamiltonian construction recipe, we end up with $L$-local Hamiltonians $H_{i,i+1,\ldots,i+L}$ obtained from the null eigenspace of $\varrho_{i,i+1,\ldots,i+L}$. Thus, for example, the following Hamiltonian
\begin{eqnarray}
& H=\sum_{i} H_{i,i+1,\ldots,i+L},
\label{Hamil}
\end{eqnarray}
where $i$ could be odd or even, provides a many-body interaction for which $|\Psi\rangle$ is a GS.

As a general example for the Hamiltonian construction, let us consider the following global state:
\begin{eqnarray}
& |\Psi\rangle_{1,\ldots,2N}=\sum_{\imath=1}^{K}\sqrt{\lambda_{\imath}}|v_{\imath}\rangle_{1,2}
|v_{\imath}\rangle_{3,4}\cdots |v_{\imath}\rangle_{2N-1,2N}.
\label{e3}
\end{eqnarray}
This specific form implies that except for $\varrho_{i,i+1}$, for odd $i$, the reduced density matrices of all other pairs of sites are separable \cite{werner}. The proof is straightforward, e.g., by using the Schmidt decomposition for the bipartite states $|v_{\imath}\rangle_{i,i+1}= \sum_{\ell} \sqrt{\zeta^{[\imath]}_{\ell}} |\ell^{[\imath]}\rangle_i |\ell^{[\imath]}\rangle_{i+1}$, for odd $i$.
Thus, the state $\varrho_{1,2}$ fully captures the pairwise (i.e, two-site) entanglement property of the whole chain. For example, if $\varrho_{1,2}$ is bound entangled, no two-site FE then can be distilled from $|\Psi\rangle$. Our examples in Sec.~\ref{exs} are all of the type of Eq.~(\ref{e3}); besides, they have zero-nullity $\varrho'_{2,3}$, hence $H$ will be many bodied.

A few remarks about the recipe for the construction of the Hamiltonian are in order here. (i) There could be many Hamiltonians (with a set of given requirements, such as symmetry, range of
interactions, and so on) for which a given $|\Psi\rangle$ is a GS. Our construction provides only one class of such Hamiltonians. (ii) By construction, our recipe implies that underlying symmetries of the state $\varrho$ would also carry over into the Hamiltonian \cite{AKMQPT,ABK}. (iii) Note that $|\Psi\rangle$ is not necessarily the unique GS of the constructed $H$. For example, each term $\ket{v_{\imath}}_{1,2} \cdots\ket{v_{\imath}}_{2N-1,2N}$ (for any $\imath$) is also a GS (one can construct other combinations which are GS). Thus, when $\varrho$ is mixed (i.e., $K>1$, which is a generic case), the GS is degenerate. This means that if we consider the system with the Hamiltonian $H$, at zero temperature the system can be everywhere in the ground eigenspace spanned by $\{|v_{\imath}\rangle_{1,2} \cdots |v_{\imath}\rangle_{2N-1,2N}\}_{\imath=1}^K$ and other possible GSs. Since this degeneracy is symmetry-driven, we can circumvent it by perturbing the system with a small symmetry-breaking term, i.e., $H'=H+\epsilon H_{\text{SB}}$, and then allowing $\epsilon\rightarrow0$ \cite{wsl-pra}. For our purposes here and for the sake of clarity, however, the working GS is assumed to be $|\Psi\rangle$ (which, for example, can be considered to be singled out by some selection rule). (iv) In most physically interesting cases, we have $L=2$ ($2$-local Hamiltonians). However, since our goal in this paper is just to provide a framework for BE QPTs, we will not be concerned with the possible appearance of many-body interactions in the Hamiltonians. Additionally, there exists a powerful perturbative gadget mechanism to construct arbitrary three-body Hamiltonians from two-body interactions \cite{gadget}. (v) Often (but not necessarily always), if the state $\varrho$ (or $|\Psi\rangle$) depends on some controllable parameter (or a set of parameters), say, $a$, so will the Hamiltonian $H(a)$. In the following, after briefly introducing GSF as the main tool for analyzing QPTs, we shall illustrate our construction through three different examples featuring BE.

\section{Order parameters}
\label{orderparameter}

Entanglement is a rather universal means of studying QPTs \cite{RMPent,sachdev}. The ground-state fidelity (GSF) \cite{GSF} has been shown to be a very powerful tool that is capable of identifying various types of QPTs, usually independent of symmetries, system-dependent details, or the nature of underlying quantum correlations; hence, it can be used as a suitable ``order parameter." Recalling that negativity,\footnote{Negativity of a bipartite state $\varrho_{AB}$ is defined as $\mathcal{N}(\varrho) =(\Vert \varrho^{T_A}\Vert_{1}-1)/2$, where $\varrho^{T_{A}}_{ab,cd}=\varrho_{cb,ad}$ (in a given basis) is the partial transposition and $\Vert A\Vert_{1}\equiv \mathrm{Tr}[\sqrt{A^{\dag}A}]$ is the trace norm \cite{neg}.} a widely used entanglement measure, fails to signal BE, and that other measures which may signal BE are usually inconclusive, GSF or its derivatives seem appropriate supplemental tools for our analysis.

The rationale behind the notion of the GSF is that, at zero temperature, the GS describes the whole characteristics of a typical quantum system. Hence, when a QPT occurs at some point $a_c$, a sudden change between the behavior of the GS slightly before ($a_c-\Delta$) and after ($a_c+\Delta$) this critical point may take place, which in principle can be captured by the following fidelity:
\begin{eqnarray}
& \mathcal{F}(a_c,\Delta) = |\langle\Psi(a_c-\Delta)|\Psi(a_c+\Delta)\rangle|.
\label{gsf}
\end{eqnarray}
The second derivative of the GSF with respect to $\Delta$, calculated at $\Delta=0$, is called the fidelity susceptibility---hereafter denoted by $\mathcal{S}$---and \textit{per se} is a powerful measure for detecting QPTs \cite{FS}. For the state $|\Psi(a)\rangle$ as in Eq.~(\ref{e3}), $\mathcal{F}$ reads
\begin{eqnarray}
\hskip -3mm &\mathcal{F}=|\sum_{\imath,\imath'=1}^K \sqrt{\lambda_{\imath} (a_-)\lambda_{\imath'} (a_+)}\braket{v_{\imath}(a_-)}{v_{\imath'}(a_+)}^N|,\hskip2mm
\label{gsf}
\end{eqnarray}
in which $a_{\pm}=a_c\pm\Delta$, and $N$ ($\Delta$) will be taken to be relatively large (small).

Here we remark that the \textit{reduced} fidelity (susceptibility)---denoting the similarity between the density matrix of a subsystem of the GS before and after the QPT point---can also be a good local order parameter for identifying symmetry-breaking~\cite{SBrF} and topological~\cite{TrF} QPTs.

There are also other, relatively less universal measures (than the GSF, but in some sense stronger than negativity or concurrence \cite{Horodeckis}) to recognize (particularly) BE associated with the TSRDM of the GS, e.g., the realignment measure (shortly, realignment) $\mathcal{N}_R$ \cite{realign}. It has been shown that if a bipartite density matrix $\varrho$ is separable, then, for the trace norm of its realignment $\varrho^R$ (defined element-wise as $\varrho^R_{ab,cd}=\varrho_{ac,bd}$), we should have $\Vert\varrho^R\Vert_{1}\leqslant1$. It has been shown that in some cases, even if $\mathcal{N}(\varrho)=0$, the positivity of $\mathcal{N}_R(\varrho)=(\Vert\varrho^R\Vert_{1}-1)/2$
can be a signature of BE \cite{realign}. Nevertheless, the nonpositivity of $\mathcal{N}_R$
does not necessarily imply separability \cite{realign}. Realignment together with negativity have been shown to set a lower bound for another widely used entanglement measure, concurrence $C(\varrho)$ \cite{Horodeckis}. More specifically, it has been proven that for an $M_1\times M_2$ ($M_1\leqslant M_2$) bipartite quantum state $\varrho$, we have $C(\varrho)\geqslant 2c_{M_1}\times\max\bigl(\mathcal{N}(\varrho),\mathcal{N}_R(\varrho) \bigr)$, where $c_{M_1}=\sqrt{2/[M_1(M_1-1)]}$ \cite{concurrence}. Moreover, realignment has been recently used in devising a scheme for directly measuring entanglement of arbitrary states \cite{cai}. Recently, a criterion more powerful than realignment for $N$-partite systems (when $N$ is even) has been proposed \cite{sr}. Based on this criterion, which provides a sufficient condition for the separability, a sparable bipartite state $\varrho$ is the one for which the relation $\mathcal N_{SR}\equiv\Vert(\varrho-\varrho_A\otimes\varrho_B)^R\Vert_1-\sqrt{(1-\mathrm{Tr}\varrho_A^2)(1-\mathrm{Tr}\varrho_B^2)}\leqslant0$ is satisfied. Here, $\varrho_A$ and $\varrho_B$ denote the reduced density matrices associated with the subsystems of $\varrho$. Since in the special examples we study in the next section $\mathcal N_R$ and $\mathcal N_{SR}$ are equal up to the multiplicative factor $1/2$, in the rest of the paper we make use of $\mathcal N_R$ as our main entanglement measure.

\section{Examples}
\label{exs}

In the following, we choose three different bipartite bound entangled states%
, and construct their corresponding Hamiltonians (all spin-one chains) as in Sec.~\ref{constH}. Our goal is to find QPTs where, for some parameter regions, a transition from BE to FE (or vice versa) or from BE to separability (or vice versa) may occur.

\subsection{Example I}
\label{HorBE}

Consider a $3\times 3$ bipartite system, consisting of two spin-1 particles (each of which can have spins $\{0,\pm1\}$, with the corresponding orthonormal states $\{|0\rangle,|1\rangle,|\bar{1}\rangle\equiv|-1\rangle\}$), whose joint density matrix $\varrho(a)$ has the following form \cite{horodec}:
 \begin{eqnarray}
& \varrho(a)= \frac{1}{1+8a}\left(
  \begin{array}{ccccccccc}
    a & 0 & 0 & 0 & a & 0 & 0 & 0 & a \\
    0 & a & 0 & 0 & 0 & 0 & 0 & 0 & 0 \\
    0 & 0 & a & 0 & 0 & 0 & 0 & 0 & 0 \\
    0 & 0 & 0 & a & 0 & 0 & 0 & 0 & 0 \\
    a & 0 & 0 & 0 & a & 0 & 0 & 0 & a \\
    0 & 0 & 0 & 0 & 0 & a & 0 & 0 & 0 \\
    0 & 0 & 0 & 0 & 0 & 0 & \frac{1+a}{2} & 0 & \frac{\sqrt{1-a^2}}{2} \\
    0 & 0 & 0 & 0 & 0 & 0 & 0 & a & 0 \\
    a & 0 & 0 & 0 & a & 0 & \frac{\sqrt{1-a^2}}{2} & 0 & \frac{1+a}{2} \\
  \end{array}
\right)\!\!,
\label{dehor}
\end{eqnarray}
where $a\in[0,1]$. This state, for all the parameter interval, has been shown to have PPT. Thus, negativity naturally fails to capture its entanglement, whereas closer inspection by the range criterion \cite{range,horodec} shows BE for $a\in(0,1)$ and separability for $a=0,1$.

To proceed with the construction of Sec.~\ref{constH}, we need the following nonzero eigenvalues and the corresponding eigenvectors of $\varrho(a)$:
\begin{eqnarray}
&\lambda_{1}=\lambda_{2}=\lambda_{3}=\lambda_{4}=\lambda_{5}=a/(1+8a),\nonumber\\
& \lambda_{6(7)}=(1+3a\mp\sqrt{1-4a+7a^2})/[2(1+8a)],
\label{eighor}
\end{eqnarray}
and
\begin{eqnarray}
\hskip -5mm &\ket{v_1}=\ket{10},~
\ket{v_2}=\ket{1\bar{1}},~\ket{v_3}=\ket{01},\nonumber\\
\hskip -5mm &\ket{v_4}=\ket{0\bar{1}},~\ket{v_5}=\ket{\bar{1}0},\\
\hskip -5mm& \ket{v_{6(7)}}=\bigl[\delta_{\pm}(\ket{11}+\ket{00})\mp \sqrt{1-a^2}\ket{\bar{1}1} +\gamma_{\mp}\ket{\bar{1}\bar{1}}\bigr]/\sqrt{Z},\nonumber
\label{vec}
\end{eqnarray}
where
\begin{eqnarray}
&\delta_{\pm}=\sqrt{1-4a+7a^2}\pm(1-3a),\nonumber\\
&\gamma_{\pm}=\sqrt{1-4a+7a^2}\pm2a,\label{dg}\\
&Z=1-a^2+2\delta_{\pm}^2+\gamma_{\mp}^2.\nonumber
\end{eqnarray}

\begin{figure}[tp]
\includegraphics[width=7.5cm,height=4.cm]{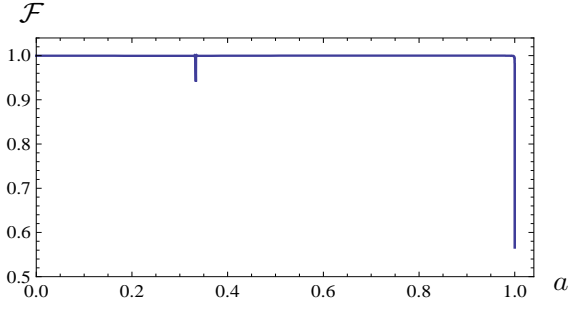}
\caption{(Color online) $\mathcal{F}$ of example I, where $N=10^8$ and $\Delta=10^{-6}$.}
\label{entfidhor}
\end{figure}

Since in this example nullity of $\rho_{2,3}$ is zero---which is also the case for the next examples---we increase the range of the interaction to three, i.e., $L=3$, and calculate the null eigenvectors of $\varrho_{1,2,3}$. This density matrix has $16$ null eigenvectors, from which only the following ones depend on $a$:
\begin{eqnarray}
&\ket{W_{\imath}}=\ket{w_1}\otimes\ket{\imath}~,~\ket{U_{\imath}}=\ket{w_2}\otimes\ket{\imath},
\label{nulls}
\end{eqnarray}
where $\imath\in\{1,0,\bar{1}\}$, $\ket{w_1}=\ket{11}+g \ket{\bar{1}1}-\ket{\bar{1}\bar{1}}$ (up to a normalization factor), $|w_2\rangle=g'|11\rangle+|00\rangle+g''|\bar{1}1\rangle -(1+g^\prime)|\bar{1}\bar{1}\rangle$ (also up to normalization), $g=\sqrt{(1-a)/(1+a)},~
g'=-2/(3+a)$, and $g''=\sqrt{1-a^2}/(3+a)$. Here, $|w_1\rangle$ and $|w_2\rangle$ are the null eigenvectors of $\varrho_{1,2}$. The other vectors in the null eigenspace can be rewritten as $\{|v_1\rangle,|v_2\rangle\} \otimes\{|0\rangle,|\bar{1}\rangle\}$, $\{|v_3\rangle,|v_4\rangle\}\otimes\{|1\rangle,|\bar{1}\rangle\}$,
and $|v_5\rangle\otimes\{|1\rangle,|0\rangle\}$. Since our aim is to study QPTs relative to the external parameter $a$, the relevant vectors in the null eigenspace of $\rho_{1,2,3}$ are the ones in Eq.~(\ref{nulls}). To have more symmetry, we construct the local Hamiltonian only from (the normalized) $\ket{W_1}$ and $\ket{W_{\bar{1}}}$, with the same coupling strengths. That is, $H_{i,i+1,i+2}=\ket{W_1}\bra{W_1}+\ket{W_{\bar{1}}}\bra{W_{\bar{1}}}$ (for odd $i$), hence $H=\sum_{i~\text{odd}} H_{i,i+1,i+2}$, or more explicitly as the following:
\begin{eqnarray}
&H=\sum_{i~\text{odd}}\bigl[8+4\{\bm{S}_{i}\cdot\bm{S}_{i+1},S^z_i
S^z_{i+1}\}_{+}-4(\bm{S}_{i}\cdot\bm{S}_{i+1})^2
\nonumber\\
&-J_1{S^z_{i}}^2+J_2{S^{z}_{i+1}}^2+J_3{S^z_i}^2
S^z_{i+1}+J_4 S^z_i {S^z_{i+1}}^2\nonumber
\\
&+g^2( {S^z_i}^2{S^z_{i+1}}^2-
S^z_{i} S^z_{i+1})+ 4g(S^z_{i+1}+S^z_{i})\nonumber\\
&+4({\{S^x,S^y\}_{i}}_{+} {\{S^x,S^y\}_{i+1}}_{+} +g(({S^z_{i}}^2-S^z_{i}){S^y_{i+1}}^2\nonumber\\
&-{S^y_{i}}^2({S^z_{i+1}}^2
+S^z_{i+1})))\bigr]{S^z_{i+2}}^2,
\label{hhor}
\end{eqnarray}
where $J_{1(2)}=4(g\pm1)~,~J_{3(4)}=-g(2\mp g)$, and $\bm{S}=(S^x,S^y,S^z)$ is the spin-1 operator, and $(A,B)_{+}\equiv AB+BA$.

\begin{figure}[tp]
\includegraphics[width=7.3cm,height=4.1cm]{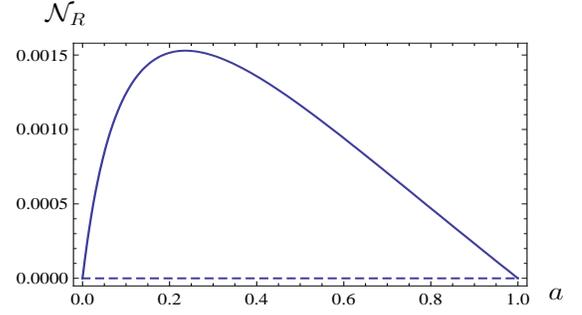}
    \caption{(Color online) $\mathcal{N}_R(\varrho)$ and $\mathcal{N}(\varrho)$ (dashed line) for example I. Note that $\mathcal{N}_R$ reaches its maximum at $a\approx0.2365$.}
\label{reahor}
\end{figure}

We now use the GSF $\mathcal{F}$ to find possible QPT points (i.e., criticalities) of this model (Fig.~\ref{entfidhor}). There are two points at which $\mathcal{F}$ shows nonanalyticity: $a_{c_1}=1$ and $a_{c_2}=1/3$.
Direct calculation shows that at both of these critical points, the behavior of the GSF for even and add $N$ is different: $\lim_{(a,\Delta)\rightarrow (1,0)}\mathcal{F}(a,\Delta)= \bigl[8+(-1)^N\bigr]/9$ and $\lim_{(a,\Delta)\rightarrow (1/3,0)}\mathcal{F}(a,\Delta)=\bigl[9+2 (-1)^N\bigr]/11$
(notice the order of this limit taking).
At $a_{c_1}$ the density matrix (\ref{dehor}) is separable; i.e., this critical point corresponds to the transition from BE to separability in the TSRDM of the GS of the system, Eq.~(\ref{dehor}), which, according to Fig.~\ref{reahor}, can also be identified by $\mathcal{N}_R$. Moreover, from Eq.~(\ref{dg}), it is simple to find that at $a_{c_1}$, $\delta_+=0$ and $\delta_-=\gamma_+$; hence, $\ket{v_6}$
and $|v_7\rangle$ become separable and maximally entangled states, respectively.
Besides, here,
$J_3=J_4=0$ and $J_1=-J_2$.
On the other hand, $\varrho$ at $a_{c_2}$ has BE. But, as clearly shown in Fig.~\ref{reahor}, this criticality cannot be captured by $\mathcal{N}_R$
(or its derivatives).

Upon closer inspection, the behavior of the GSF at very large (small) $N$ ($\Delta$) reveals another possibly critical point in this model at $a_{c_3}=0$.
Since $\varrho$ becomes separable at this point, this criticality corresponds to the transition from separability to BE in the TSRDM of the GS.


\subsection{Example II}

Consider the following family of three-parameter $3\times 3$ states:
\begin{eqnarray}
&\chi(\alpha,\beta,a)=\frac{1}{9}(1-\alpha-\beta-a)\openone\otimes\openone+\alpha P_{00}\nonumber\\ &+\frac{\beta}{2}(P_{10}+P_{20})+\frac{a}{3}(P_{01}+P_{11}+P_{21}),\label{mum}
\end{eqnarray}
where $P_{mm'}=(U_{mm'}\otimes\openone)|\Phi^+\rangle\langle\Phi^+|(U^\dagger_{mm'}\otimes\openone)$, $U_{mm'}=\sum_{m''=0}^{2}e^{\frac{2\pi i}{3}m m''}|m''\rangle\langle(m'+m'')~\text{mod}~3|$ ($m,m'\in\{0,1,2\}$), and $\ket{\Phi^+}=\sum_{l=0}^2\ket{ll}/\sqrt{3}$. The entanglement properties of $\chi(\alpha,\beta,a)$ have been studied, by using a geometric entanglement witness, in Ref.~\cite{bert}, and $\chi$ appears to have PPT BE for $a\neq 0$.
\begin{figure}[tp]
\includegraphics[width=7.1cm,height=4.3cm]{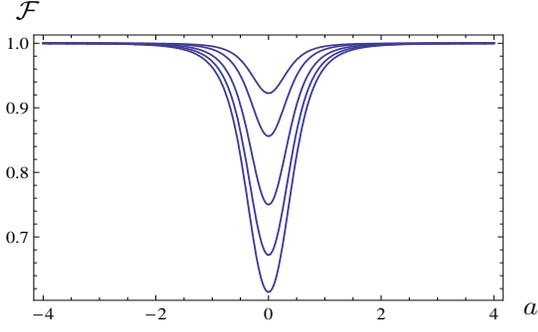}
\caption{(Color online) $\mathcal{F}$ of example II [Eq.~(\ref{simstate})] vs $a$, for $\Delta=10^{-5}$ and $N=10^9,2\times 10^9,4\times 10^9,6\times 10^9,$ and $8\times 10^9$ from top to bottom.}
\label{simgsf}
\end{figure}

To enable the method introduced earlier for the construction of the Hamiltonian $H$, it is required that the TSRDM $\varrho$ have at least one null eigenvector that depends on a (set of) parameter (parameters). Note, however, that since the eigenvectors of $\chi(\alpha,\beta,a)$ are parameter independent, this state cannot be used as $\varrho$ for our mechanism. Instead, we consider the following modification:
\begin{eqnarray}
\hskip-5mm&\varrho= p~\bigl[q~\chi^{T_A}+
\frac{(1-q)}{9}~\openone\otimes \openone\bigr]+(1-p)~\ket{\Phi^+}\bra{\Phi^+}. \label{simstate}
\end{eqnarray}
The parameters must be chosen such that the positivity of the state $\varrho$ is guaranteed.
Also, by suitably choosing the parameter $q$, we can make some of the eigenvalues of $\varrho$ vanish. In addition, introducing the parameter $p$ is another way by which we can adjust the amount of entanglement of $\varrho$ so that a criticality occurs in our desired point.

Because of the special form of $\chi$, the eigenvectors of $\varrho$, apart from a dependence on the sign of $pq$, do not depend on the absolute values of $p$ and $q$. This means that the parent Hamiltonian $H$ has a continuous degeneracy with respect to the parameters $p$ and $q$. In the rest of the discussion, we assume $0<p<1$.

There are two proper choices for the parameter $q$: $q^\pm=2/[2 \alpha +2 \beta -a \pm3 \sqrt{(\beta -2 \alpha )^2+a ^2}]$, either of which can make three eigenvalues of $\varrho$ vanish. For $q^\pm$, the sets of three null eigenvectors $v^\pm=\{v^\pm_{\imath}\}_{\imath=1}^3$ are
\begin{eqnarray}
\label{Vs}
&|v_1^\pm\rangle=(b^\pm_{1}\ket{10}+\ket{01})/\sqrt{1+(b^\pm_{1})^2},\nonumber\\
&|v_2^\pm\rangle=(b^\pm_{2}\ket{1\bar{1}}+\ket{\bar{1}1})/\sqrt{1+(b^\pm_{2})^2},\\
&|v_3^\pm\rangle=(b^\pm_{1}\ket{0\bar{1}}+\ket{\bar{1}0})/\sqrt{1+(b^\pm_{1})^2},\nonumber
\end{eqnarray}
in which $b^\pm_1=[a\mp\text{sgn}(q)\sqrt{(\beta -2 \alpha )^2+a^2}]/(2 \alpha -\beta)$ and $b^\pm_2=b^\pm_1(a\to-a)$. Note that when we choose $q^\pm$, the set $\{v^\mp\}$ corresponds to nonzero eigenvalues $\lambda_{1,2,3}=p~\{q~[a -2 (\alpha +\beta )]+2\pm3q~\text{sgn}(q)\sqrt{(\beta -2 \alpha )^2+a^2}\}/18$. The other eigenvalues of $\varrho$, which we require to be nonzero (otherwise the degeneracy of the parent Hamiltonian would increase), are $\lambda_4=\{[2q(\alpha +\beta )-q a -8]p+9\}/9$ and $\lambda_{5}=\lambda_{6}= p[2 q(\alpha +\beta )-q a +1]/9$, corresponding to the following eigenvectors:
\begin{eqnarray}
&|v_4\rangle=(|11\rangle+|00\rangle+|\bar{1}\bar{1}\rangle)/\sqrt{3},\nonumber\\
&|v_5\rangle= (-|11\rangle+|\bar{1}\bar{1}\rangle)/\sqrt{2},\label{non-z}\\
&|v_6\rangle=(-|11\rangle+2|00\rangle-|\bar{1}\bar{1}\rangle)/\sqrt{6}.\nonumber
\end{eqnarray}
\begin{figure}[tp]
\vskip.35cm
\includegraphics[width=7.2cm,height=4.cm]{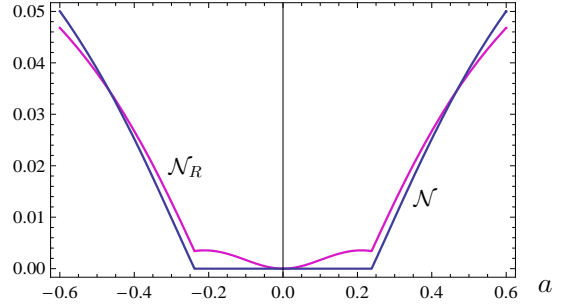}
\caption{(Color online) $\mathcal{N}$ and $\mathcal{N}_R$ of example II-a vs $a$.}
\label{simneg}
\end{figure}
Also note that, irrespective of the values of the parameters in Eq.~(\ref{simstate}), $\varrho'_{i,i+1}$ (the TSRDM of sites $i$ and $i+1$), for even $i$
has an empty null eigenspace. Thus, as explained in the previous example, we consider $\rho_{1,2,3}$ for the construction of the Hamiltonian $H$. The null eigenspace of $\rho_{1,2,3}$ is spanned by $13$ vectors, from which only the following six are parameter dependent:
\begin{eqnarray}
&\ket{W_{1,0}^\pm}=|v_1^\pm\rangle\otimes\{|1\rangle,|0\rangle\},\nonumber\\
&\ket{U_{1,\bar{1}}^\pm}=|v_2^\pm\rangle\otimes\{|1\rangle,|\bar{1}\rangle\}, \\
&\ket{V_{0,\bar{1}}^\pm}=|v_3^\pm\rangle\otimes\{|0\rangle,|\bar{1}\rangle\},\nonumber
\label{simhvect}
\end{eqnarray}
where $\pm$ correspond to $q^{\pm}$. The other null eigenvectors are $\{|0\bar{1}\rangle,|\bar{1}0\rangle\}\otimes|1\rangle,~\{|1\bar{1}\rangle,|\bar{1}1\rangle\}
\otimes|0\rangle,~\{|10\rangle,|01\rangle\}\otimes|\bar{1}\rangle$ and $|v_5\rangle\otimes|0\rangle$. For odd $i$, we write the local Hamiltonian as $H_{i,i+1,i+2}=\ket{W_0^\pm}\bra{W_0^\pm}+\ket{V_0^\pm}\bra{V_0^\pm}$
and obtain the global Hamiltonian
\begin{eqnarray}
&\hskip-3mm H=\sum_{i~\text{odd}}\bigl[J_1\bigl(\bm{S}_i \cdot\bm{S}_{i+1}+\{\bm{S}_i \cdot\bm{S}_{i+1},S^z_i S^z_{i+1}\}_+ - S^z_i S^z_{i+1}\bigr)\hskip-1mm\nonumber\\
&~+J_2{S^z_i}^2{S^z_{i+1}}^2+J_3\bigl({S^z_i}^2 S^z_{i+1}-S^z_i{S^z_{i+1}}^2+S^z_i-S^z_{i+1}\bigr)\hskip-1mm\nonumber\\
&~+{S^z_i}^2+{S^z_{i+1}}^2\bigr]{(\openone-{S^z}^2)_{i+2}},\label{SB'sH}
\end{eqnarray}
where
\begin{eqnarray}
&J_1=2\mu_1^\pm\nu_1^\pm,~J_2=-2(\mu_1^\pm+\nu_1^\pm)^2,~J_3=(\mu_1^\pm)^2-(\nu_1^\pm)^2,\nonumber\\
&\mu^\pm_1=b^\pm_1/\sqrt{1+(b^\pm_1)^2},~~\nu^\pm_1=\sqrt{1-(\mu^\pm_1)^2}.
\end{eqnarray}

$H$ represents a three-parameter family of Hamiltonians whose parameters are determined so that the positivity of $\varrho$ is guaranteed. In the following, we study the entanglement properties of the GS of this Hamiltonian in the vicinity of its critical points, for two special cases.

\begin{figure}[tp]
\includegraphics[width=7cm,height=4.3cm]{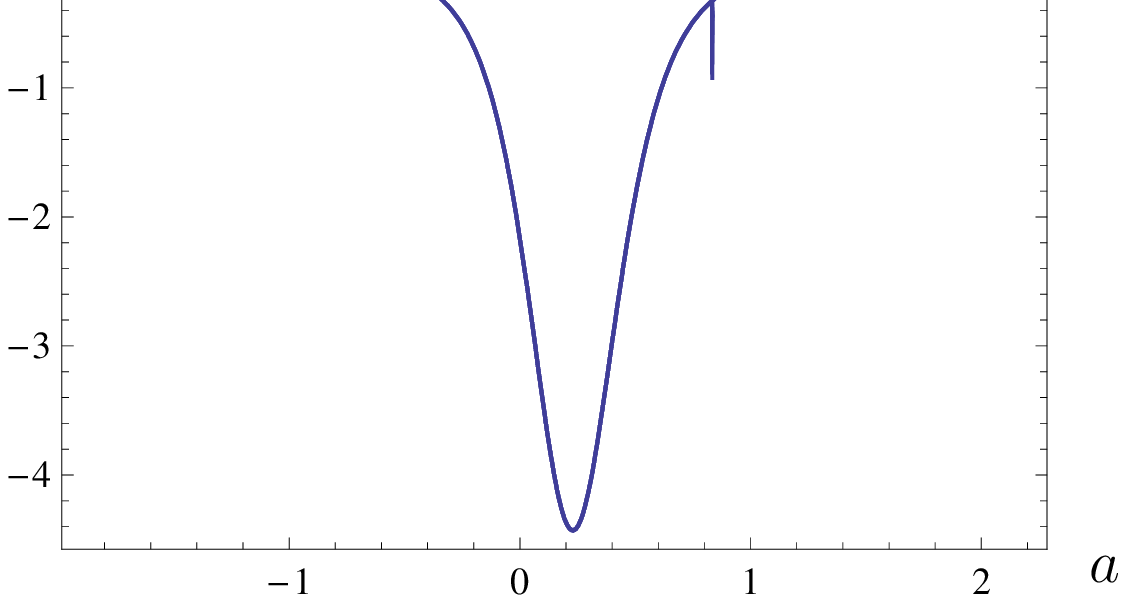}
\caption{(Color online) $\mathcal{S}/N$ of example II-b vs $a$.}
\label{FS1}
\end{figure}

\subsubsection{Example II-a}
When $\alpha=(6-b)/21,~\beta=-2b/21$, and $a=(5-2b)/7$ [i.e., $\alpha=(1+a)/6$ and $\beta=(-5+7a)/21$], $\chi(\alpha,\beta,a)$
is reduced to another bound entangled state \cite{horalpha}
$\widetilde{\chi}=[2\ket{\Phi^+}\bra{\Phi^+}+b\sigma_+ +(5-b)\sigma_-]/7$, in which  $\sigma_{+}=(\ket{10}\bra{10}+\ket{0\bar{1}}\bra{0\bar{1}}+\ket{\bar{1}1}\bra{\bar{1}1})/3$, and $\sigma_-$
is the swap of $\sigma_+$. 
For $b\in(3,4]$, $\widetilde{\chi}$ has BE. Let us assume that $q=q^+$ and $p=3/4$.
The parent Hamiltonian then will be $H$ in Eq.~(\ref{SB'sH}) with superscript $+$. The GS (\ref{e3}) is written with the set $v^-$ and its
corresponding eigenvalues, $\{\lambda_1,\lambda_2,\lambda_3\}|_{q=q^+}$, plus eigenvalues $\{\lambda_4,\lambda_5,\lambda_6\}|_{q=q^+}$ and
eigenvectors of Eqs.~(\ref{non-z}). In this way, the GSF (\ref{gsf}) can be obtained readily. Figure~\ref{simgsf} shows the finite-size
behavior of $\mathcal{F}$ vs $a$, for $\Delta=10^{-5}$ and different $N$. As is seen, $\mathcal{F}$ has a minimum at $a_c=0$ for all values of $N$.
When $N$ increases, $\mathcal{F}$ decreases; however, the rate of decreasing at $a_c=0$ ($\forall N$) is maximum. This behavior can be a sign of
criticality in the system. At this point, $b_1^\pm=b_2^\pm$ and negativities of $|v_1^\pm\rangle$, $|v_2^\pm\rangle$, and $|v_3^\pm\rangle$ attain
their maximum values. On the other hand, as seen in Fig.~\ref{simneg}, $a_c=0$ is in the region where $\mathcal{N}(\varrho)$ vanishes; hence,
$\mathcal{N}$ fails to signal criticality in the system. From the positivity of $\mathcal{N}_R (\varrho)$ around $a_c$, it is evident that $a=0$ is
the point where the bipartite entanglement of the GS experiences transition from BE to separability---$\mathcal{N}_R$ detects this critical point.
Furthermore, despite the nonanalyticity of the first derivative of $\mathcal{N}_R$ at points $\varrho$ goes from the FE to the BE
region, there is no critical behavior in $\mathcal{F}$ at these points. This implies that the transition from BE to FE cannot be responsible for QPT in this system.
\begin{figure}[tp]
\vskip1.2mm
\includegraphics[width=6.7cm,height=4.1cm]{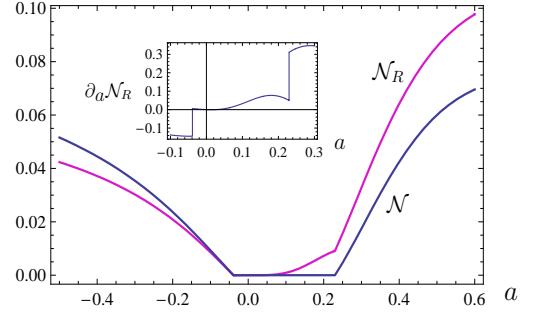}
\caption{(Color online) $\mathcal{N}$ and $\mathcal{N}_R$ of example II-b vs $a$. Inset: The first derivative of $\mathcal{N}_R$ with respect to $a$.}
\label{SS10}
\end{figure}

\subsubsection{Example II-b}
Suppose that $\alpha=(-1+3a)/6$, $\beta=(1+3a)/7$, $q=q^-$, and $p=0.76027256$.
According to
our previous explanations, the Hamiltonian $H$ (the GS $|\Psi\rangle$) is constructed from the set $v^-$ [$v^+$ and its corresponding eigenvalues
$\{\lambda_{1},\lambda_{2}, \lambda_{3}\}|_{q=q^-}$, together with the eigenvalues $\{\lambda_{4},\lambda_{5},\lambda_{6}\}|_{q=q^-}$, and the
eigenvectors (\ref{non-z})]. Thus, the global Hamiltonian of this model will be $H$ in Eq.~(\ref{SB'sH}) with superscript $-$. In Fig.~\ref{FS1}, $\mathcal{S}/N$ has been plotted in terms of $a$ for different values of $N$. There exist two points at which this system may exhibit criticality. At $a_{c_1}=5/6$, $\mathcal{S}/N$ becomes nonanalytic. Indeed, for arbitrary values of $p$ at this point, we have $\lim_{(a,\Delta)\rightarrow (5/6,0)}\mathcal{F}(a,\Delta,p)= \{11+10 [-1+(-1)^N] p\}/11$. At $a_{c_1}$, since $2\alpha=\beta$, then $b_{1(2)}^\pm\to\infty$. Therefore, vectors $|v^\pm\rangle$ in Eq.~(\ref{Vs}) at $a_{c_1}$ are completely factorized. Furthermore, one
should note that at this point, in the global Hamiltonian $H$, $\mu_{1(2)}^+\to 1$ and then $\nu^+_{1(2)}$ goes to zero. As clearly shown in Fig.~\ref{SS10}, this probable QPT occurs when $\varrho$ is completely free entangled, and of course negativity does not
show it (note that this result is due to the existence of a trace norm in the definition of negativity since the first derivative of one of the eigenvalues of $\varrho^{T_A}$ has a discontinuity at $\eta_{c_1}$ and therefore can signal it). Another possible QPT point of this system is $a_{c_2}\approx 0.229650$. As the behavior of $\mathcal{S}/N$ in Fig.~\ref{FS1} shows, the rate of the drop in $\mathcal{F}$ at this point is maximum, which may signal a QPT in the system. The interesting fact is that this critical point corresponds to a change in the type of entanglement of the TSRDM of the GS. In fact, since $a_{c_{2}}$ is the initial point of the range at which $\mathcal N$ is zero and $\mathcal N_R$ is positive (see Fig.~\ref{SS10}), it represents a transition from FE to BE in the two-site states. As Fig.~\ref{SS10} shows, the first derivative of $\mathcal N_R$ at $a_{c_2}$ is discontinuous, implying that it detects this criticality.

\subsection{Example III}

As the final example, consider the following $3\times3$ density matrix:
\begin{eqnarray}
\label{statesah}
& \varrho(a)=\frac{1}{20 g b^2}\left(
  \begin{array}{ccccccccc}
    \gamma & \omega & 0 & 0 & \sigma & 0 & 0 & 0 & \mu \\
    \omega & 2\nu &
    0
     & 0 & 0 & 0 & 0 & 0 & \beta \\
    0 & 0 & \nu & 0 & 0 & 0 & 0 & 0 & 0 \\
    0 & 0 & 0 & \nu & 0 & 0 & 0 & 0 & 0 \\
    \sigma & 0 & 0 & 0 & b^4 & 0 & 0 & 0 & \eta \\
    0 & 0 & 0 & 0 & 0 & \nu & 0 & 0 & 0 \\
    0 & 0 & 0 & 0 & 0 & 0 & \nu & 0 & 0 \\
    0 & 0 & 0 & 0 & 0 & 0 & 0 & \nu & 0 \\
    \mu & \beta & 0 & 0 & \eta & 0 & 0 & 0 & \epsilon \\
  \end{array}
\right),
\end{eqnarray}
in which
\begin{eqnarray}
\label{sahparameter}
&g=1+a+a^2~,~b=\sqrt{a+g+1}~,~\gamma=(1+g)^2,\nonumber\\
&\omega=2gb~,~\sigma=b^2(g-1)~,~\mu=(1+a)(1+3g+a),~~\\
&\nu =2gb^2~,~\beta=2gb(1+a)~,~\eta=-b^2a,~\epsilon=(2g+a)^2.\nonumber
\end{eqnarray}
\begin{figure}[tp]
\vskip3.6mm
    \includegraphics[width=7.5cm,height=4cm]{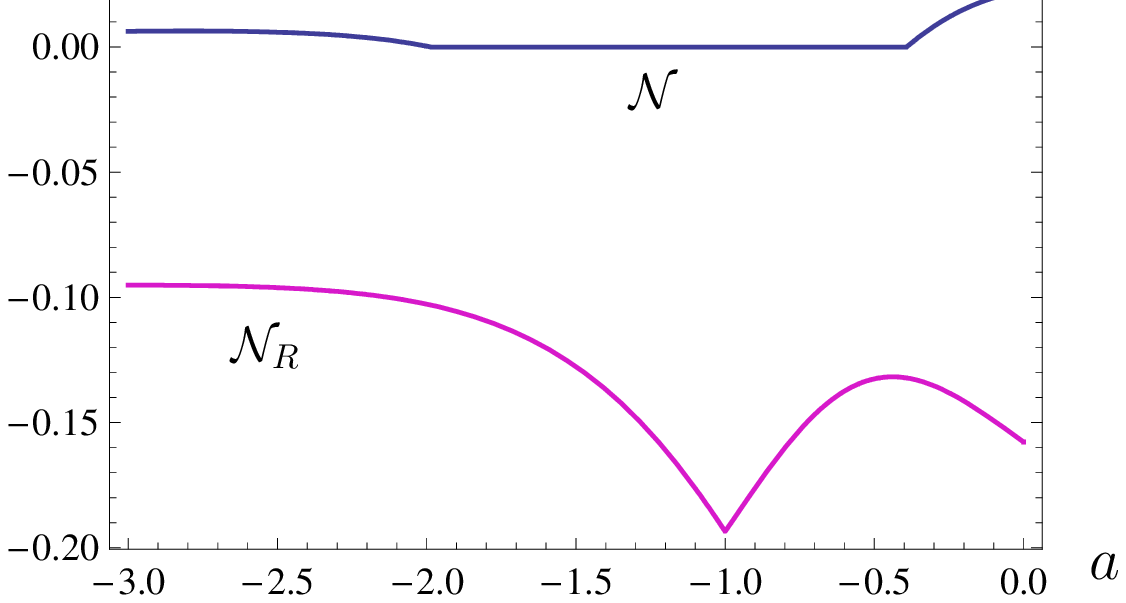}
    \caption{(Color online) $\mathcal{N}$ and $\mathcal{N}_R$ of example III vs $a$.}
\label{sa-realign}
\end{figure}
After some straightforward algebra, we find the following nonzero eigenvalues and their corresponding eigenvectors:
 \begin{eqnarray}
 \label{saheig}
&\lambda_1=3/10,~\lambda_{2}=\cdots=\lambda_{8}=1/10,\nonumber\\
&\ket{v_{1(2)}}=[\ket{11}\pm b\ket{10}+(1+a)\ket{\bar{1}\bar{1}}]/(1+a),\nonumber\\
&\ket{v_3}=\bigl[a(a+1)\ket{11}+b^2\ket{00}-a\ket{\bar{1}\bar{1}}\bigr]/b^2,\\
&\ket{v_4}=\ket{1\bar{1}},~\ket{v_5}=\ket{01},~\ket{v_6}=\ket{0\bar{1}},\nonumber\\
&\ket{v_7}=\ket{\bar{1}1},~\ket{v_8}=\ket{\bar{1}0}.\nonumber
\end{eqnarray}
For the same reason as in the previous examples, i.e., the nullity of $\varrho_{2,3}$ constructed from Eq.~(\ref{e3}) is zero, we work with $\varrho_{1,2,3}$ to construct the Hamiltonian. The null eigenspace of $\rho_{1,2,3}$ is spanned by
\begin{eqnarray}
\hskip-6mm&\ket{W_{\imath}}=\bigl[-(1+a)\ket{11}+a\ket{00} +\ket{\bar{1}\bar{1}}\bigr]\otimes\ket{\imath},\nonumber\\
\hskip-6mm&\ket{W'}=\bigl[-(\frac{1+a}{a})\ket{11}+\ket{00} +\frac{1}{a}\ket{\bar{1}\bar{1}}\bigr]\otimes|0\rangle,\label{sahnull}\\
\hskip-6mm&\ket{U}=(a'|11\rangle+|\bar{1}\bar{1}\rangle)\otimes|0\rangle,\nonumber
\end{eqnarray}
together with $\{|\bar{1}0\rangle,|\bar{1}1\rangle\}\otimes\{|1\rangle,|0\rangle\}$, $\{|01\rangle,|0\bar{1}\rangle\}\otimes\{|1\rangle,|\bar{1}\rangle\}$, $|1\bar{1}\rangle\otimes\{|0\rangle,|\bar{1}\rangle\}$, and $|100\rangle$, where
$\imath\in\{1,\bar{1}\}$ and $a'=1/(1+a)$. We consider the local Hamiltonian as $H_{i,i+1,i+2}=\ket{U}\bra{U}$, for odd $i$, hence
\begin{eqnarray}
&H=\sum_{i~\text{odd}}\bigl\{J_1[2+\{\bm{S}_{i}\cdot\bm{S}_{i+1},S^z_i
S^z_{i+1}\}_{+}-(\bm{S}_{i}\cdot\bm{S}_{i+1})^2
\nonumber\\
&-{S^z_{i}}^2-{S^{z}_{i+1}}^2+{\{S^x,S^y\}_{i}}_{+} {\{S^x,S^y\}_{i+1}}_{+}]\nonumber\\
&+J_2({S^z_i}^2{S^z_{i+1}}^2+S^z_{i} S^z_{i+1})+ J_3({S^z_i}^2{S^z_{i+1}}+ {S^z_i}{S^z_{i+1}}^2)\bigr\}\nonumber\\&\times(\openone-{S^z}^2)_{i+2},
\end{eqnarray}
in which
\begin{eqnarray}
&J_1=-4 a'a'',~J_2=(a'+a'')^2,~J_3=(2a'^2-1),
\end{eqnarray}
and $a''=\sqrt{1-a'^2}$.
\begin{figure}[tp]
    \includegraphics[width=7.1cm,height=4.3cm]{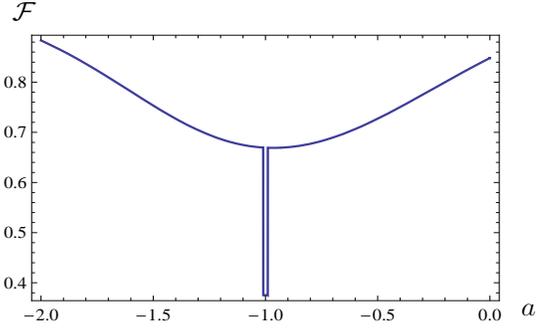}
    \caption{(Color online) $\mathcal{F}$ of example III vs $a$, for $N=10001$ and $\Delta=0.01$.}
    \label{sa-realign2}
\end{figure}

Figure~\ref{sa-realign} shows the behavior of negativity $\mathcal{N}$ and realignment $\mathcal{N}_R$ vs $a$. At $a\in(-2,-0.4]$,
$\mathcal{N}$ remains zero, whereas $\mathcal{N}_{R}$
is negative (with a nonanalyticity at $a_c=-1$).
As explained before, although for a separable state $\mathcal{N}_R$ is nonpositive, the converse is not necessarily true. We could not conclude whether any BE can exist in this specific region of $a$ because we do not know of any other measure stronger than
$\mathcal{N}_{SR}$ (whose behavior---as explained in Sec.~\ref{orderparameter}---is the same as $\mathcal N_R$ up to the multiplicative factor $1/2$) for detecting BE. We, nonetheless, show that a criticality occurs in this region exactly at the nonanalyticity of $\mathcal{N}_R$.
Figure~\ref{sa-realign2} depicts the GSF (\ref{gsf}) of this model, showing a QPT at $a_c=-1$. At this point, $\lim_{(a,\Delta)\rightarrow (-1,0)}\mathcal{F}(a,\Delta)= [2(-1)^N+3]/5$, and nonanalyticity at $a_c$ appears for odd $N$s.

Finally, we remark that we investigated the behavior of the single- and two-site fidelity (susceptibility) of our examples around their critical points. The results show that in most of the cases, the total behavior of these local order parameters are similar to that of their global counterparts, hence they are also capable of detecting quantum critical points associated with our models. For example, the single-site fidelity in example I shows a sharp drop at $a_{c_1}$ and the two-site fidelity (susceptibility) in example II-a (example II-b) can identify all the QPT points. Nevertheless, there exist some critical points, e.g., $a_{c_2}$ in example I, for which these local measures do not herald quantum criticality. 

\section{Summary and discussion}
\label{conclusion}

Here we have investigated the role of bound entanglement in quantum phase transitions. To this aim, we have presented a method to construct quantum spin-chain Hamiltonians in which a quantum phase transition can be accompanied by a change in the type of entanglement (of the two-site reduced density matrices of the corresponding ground states) from bound to free or separability. This method \textit{per se} is fairly general and can be applied to various scenarios.  Given a form for the desired two-site reduced density matrices (with their engineered entanglement properties), by utilizing their spectral properties, we have outlined a reverse construction for a compatible pure state for the whole chain. Additionally, we have suggested a method to obtain a class of Hamiltonians for which the constructed pure state is a ground state. At zero temperature, then, by varying the Hamiltonian parameters (carried over from the very parameter dependence of the given reduced density matrices), a desired quantum phase transition at some engineered points can take place. To be able to detect possibly relevant criticalities (due to the underlying entanglement-type changes), we have used the ground-state fidelity and the realignment criterion.

One direction for extending our results is to delve into a possible connection between our construction of Hamiltonians and the methods of quantum inverse scattering \cite{Faddeev}. Although these methods have clear distinctions, there might be a rich common framework through which one can theoretically engineer quantum models with prescribed entanglement properties for ground states.

Overall, our study may hopefully spur further interest in bound entanglement and the role it may play in quantum phase transitions or other areas in which a potential application or advantage could be anticipated. For example, quantum communications \cite{chdis,qcom,qcom-science} and controllable generation of entanglement in engineered physical systems \cite{entcont} could be possible beneficiaries.

\section{Acknowledgments}

Comments by H. Johannesson, V. Karimipour, and D. A. Lidar are gratefully acknowledged.


\end{document}